# Efficiency Evaluation Metrics for Wireless Intelligent Sensors Applications

**Saad Chakkor**
University of Abdelmalek Essaâdi, Faculty of Sciences, Department of Physics,
Communication and Detection Systems Laboratory, Tetouan, Morocco
Email: saadchakkor@gmail.com

**El Ahmadi Cheikh, Mostafa Baghouri, Abderrahmane Hajraoui**
University of Abdelmalek Essaâdi, Faculty of Sciences, Department of Physics,
Communication and Detection Systems Laboratory, Tetouan, Morocco
Email: {elahmadi13, baghouri.mostafa}@gmail.com, ad_hajraoui@hotmail.com

***Abstract*— The metrology field has been progressed with the appearance of the wireless intelligent sensor systems providing more capabilities such as signal processing, remote multi-sensing fusion etc. This kind of devices is rapidly making their way into medical and industrial monitoring, collision avoidance, traffic control, automotive and others applications. However, numerous design challenges for wireless intelligent sensors systems are imposed to overcome the physical limitations in data traffic, such as system noise, real time communication, signal attenuation, response dynamics, power consumption, and effective conversion rates etc, especially for applications requiring specific performances. This paper analyzes the performance metrics of the mentioned sensing devices systems which stands for superior measurement, more accuracy and reliability. Study findings prescribe researchers, developers/ engineers and users to realizing an optimal sensing motes design strategy that offers operational advantages which can offer cost-effective solutions for an application.

## I. INTRODUCTION

Wireless technologies have made significant progress in recent years, allowing many applications in addition to traditional voice communications and the transmission of high-speed data with sophisticated mobile devices and smart objects. In fact, they also changed the field of metrology especially the sensor networks and the smart sensors. The establishment of an intelligent sensor system requires the insertion of wireless communication which has changed the world of telecommunications. It can be used in many situations where mobility is essential and the wires are not practical. Today, the emergence of radio frequency wireless technologies suggests that the expensive wiring can be reduced or eliminated. Various technologies have emerged providing communication differently. This difference lies in the quality of service and in some constraints related on the application and it environment. The main constraints to be overcome in choosing a wireless technology revolve around the following conditions [1], [2]:

- Range
- Reliability
- Bandwidth
- conformity (standards)
- Security
- Cost
- Energy consumption
- Speed and transmission type (synchronous, asynchronous)
- Network architecture (topology)
- Environnement (noise, obstacles, weather, hypsometry)

In this work, we studies using a comparative analysis, the different parameters which influence the performance and quality of a wireless communication system based on intelligent sensors taking into our consideration the cost and the application requirements. We can classify the requirements of applications using smart sensors into three main categories as shown in table 1.

Table 1. Needs based applications

| Types of application | Specifications and Needs |
|---|---|
| Environmental monitoring | ▪ Measurement and regular sending<br>▪ Few data<br>▪ Long battery life<br>▪ Permanent connection |
| Event detection | ▪ Alert message<br>▪ Priority<br>▪ Confirmation status<br>▪ Few data<br>▪ Permanent connection |
| Tracking | ▪ Mobility<br>▪ Few data<br>▪ Localization<br>▪ Permanent connection |

Section II present the related works realized in this context. However section III summarize the main constraints of intelligent sensor systems. Moreover, in section IV the performance indicators are defined and discussed, in some cases, compared with those derived from a theoretical as well as simulation analysis results. Section V provides performance comparison of the popular sensor motes. Finally, section VI concludes the paper and outlines some future directions of research.

## II. Related Work

In the related work, many research studies in [3-8] have been focused on wireless sensor networks to improve communication protocols in order to solve the energy constraint, to increase the level of security and precision and to expand autonomy for accuracy, feasibility and profitability reasons. On the other side, the field of intelligent sensors remains fertile and opens its doors to research and innovation, it is a true technological challenge in so far as the topology and the infrastructure of the systems based on intelligent sensors are greatly different compared to wireless sensor networks, particularly in terms of size (number of nodes) and routing. In fact, to preserve the quality of these networks, it is very difficult even inconceivable to replace regularly the faulty nodes, which would result in a high cost of maintenance. The concept of energy efficiency appears therefore in communication protocols, [5-9]. Thus, it is very useful to search the optimization of data routing and to limit unnecessary data sending and the collisions [6], [9]. The aim challenge for intelligent sensors systems is to overcome the physical limitations in data traffic such as system noise, signal attenuation, response dynamics, power consumption, and effective conversion rates etc… This paper emphasis on the performance indicators for wireless protocols which stands for superior measurement, more accuracy and reliability. The object of this study is for realizing an advanced intelligent sensors strategy that offers many system engineering and operational advantages which can offer cost-effective solutions for an application.

## III. NEW CONSTRAINTS OF INTELLIGENT SENSOR SYSTEMS

An intelligent sensor is an electronic device for taking measurements of a physical quantity as an electrical signal, it intelligence lie in the ability to check the correct execution of a metrology algorithm, in remote configurability, in its functions relating to the safety, diagnosis, control and communication. The intelligent sensor can be seen consisting of two parts [10-13]:

1) A measuring chain controlled by microcontroller
2) A bidirectional communication interface with the network, providing the connection of the sensor to a central computer

The communication part reflects all the information collected by an intelligent sensor and allows the user to configure the sensor for operation. It is therefore absolutely essential that this interface be robust and reliable. Figure 1 illustrates the intelligent sensor with its wireless communicating component. A variety of communication interfaces (wireless modules) is available, but not all sensors support these interfaces. The designer must select an interface that provides the best integration of the sensor with the others components of the system taking in our account the costs and the constraints of reliability required for a particular application. There are others solutions to collect remote measurements such mobile and satellite communications. The main problems related to the quality of communications are: attenuation problems (distance, obstacles, rain ...), interference and multipath. The realization of the systems based on smart sensors dedicated to the applications mentioned in section I, requires the techniques and the protocols that take into account the following constraints [3]:

- The nodes are deployed in high numbers
- At any time, the nodes may be faulty or inhibited
- The topology changes very frequently
- The communication is broadcast

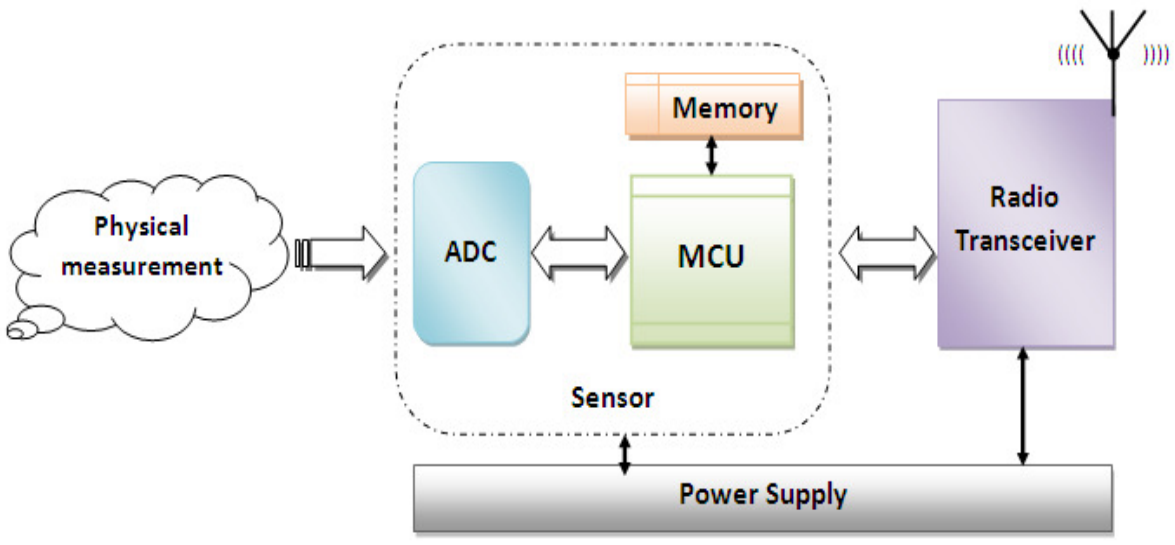

Fig 1. Block diagram of an intelligent sensor communication

In other way, the intelligent sensors have benefits like ad-hoc topology which is easy to deploy, low cost, sensing capabilities in difficult environment. Regarding their challenges they are limited in energy, in computing capacity, in size of sensors and in memory. They need also security in term of data confidentiality and technology. For ad-hoc networks, energy consumption was considered as an aim factor but not essential because energy resources can be replaced by the user. These networks are more focused on the QoS than the energy consumption. Contrariwise, in sensor networks, the transmission time and energy consumption are two important performance metrics since generally the sensors are deployed in a vast inaccessible areas.

## IV. PERFORMANCE INDICATORS

Intelligents wireless sensor networks are different from the traditional communication networks, and therefore different performance measures may also be required to evaluate them. Among them are the QoS metrics. Some applications in WSNs have real-time properties. These applications may have QoS requirements such as delay, loss ratio, and bandwidth. This section describes the different criteria to evaluate the performance of a wireless communication for intelligent sensors systems.

### A. *Network Size and topology*

The size and the topology of a sensors network can be adopted according to differents application requirements such as data packets size during traffic, transmission protocols implemented, interferences and dimensions of the monitored area. Table 2 summarizes the main differences in size and in topology between the mentioned protocols.

data payload size $N_{data}$ and it's not proportional to the maximum data rate.

Table 2. Network size and basic topology

| Protocol | Bluetooth | UWB | ZigBee | Wi-Fi | Wi-Max | GPRS |
|---|---|---|---|---|---|---|
| Basic Cell | Piconet | Piconet | Star | BSS | Single Cell | Single Cell |
| Extension of the Basic Cell | Scatternet | P2P | Cluster tree, Mesh | ESS | PTMP, PTCM, Mesh | Cell System |
| Max number of Cell Nodes | 8 | 236 | > 65000 | 2007 | 1600 | 1000 |

### B. Transmission Time

The transmission time depends on the data rate, the message size, and the distance between two nodes. The formula of transmission time in (µs) can be described as follows:

$$T_{tx} = \left( N_{data} + \left( \frac{N_{data}}{N_{maxPld}} \times N_{ovhd} \right) \right) \times T_{bit} + T_{prop} \qquad (1)$$

$N_{data}$ the data size
$N_{maxPld}$ the maximum payload size
$N_{ovhd}$ the overhead size
$T_{bit}$ the bit time
$T_{prop}$ the propagation time between two nodes to be neglected in this paper

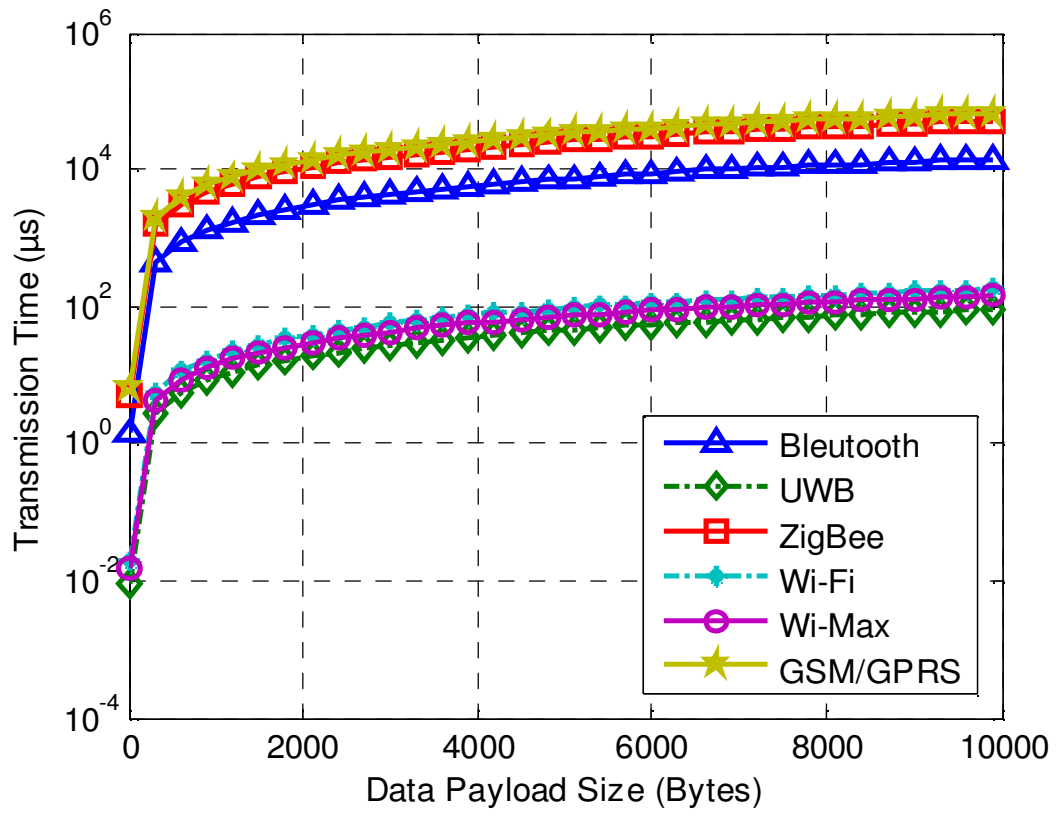

Fig 1. Comparison of transmission time relative to the data size

From the figure 2, it is noted that the transmission time for the GSM/GPRS is longer than the others, due to its low data rate (168 Kb/s) and its long range reasons, while UWB requires less transmission time compared to the others because of its important data rate. It clearly shows that the required transmission time is proportional to the

The typical parameters of the different wireless protocols used to evaluate the time of transmission are given in Table 3.

Table 3. Typical parameters of wireless protocols

| Protocol | Max data rate (Mbit/s) | Bit time (µs) | Max data payload (bytes) | Max overhead (bytes) | Coding efficiency+ (%) |
|---|---|---|---|---|---|
| **Bluetooth** | 0.72 | 1.39 | 339 (DH5) | 158/8 | 94.41 |
| **UWB** | 110 | 0.009 | 2044 | 42 | 97.94 |
| **ZigBee** | 0.25 | 4 | 102 | 31 | 76.52 |
| **Wi-Fi** | 54 | 0.0185 | 2312 | 58 | 97.18 |
| **Wi-Max** | 70 | 0.0143 | 2700 | 40 | 98.54 |
| **GPRS** | 0.168 | 5.95 | 1500* | 52* | 80.86 |

+ Where the data is 10 Kbytes. * For TCP/IP Protocol

### C. Transmission power and range

In wireless transmissions, the relationship between the received power and the transmitted power is given by the Friis equation as follows [1], [33], [36-40]:

$$\frac{P_r}{P_t} = G_t G_r \left( \frac{\lambda}{4\pi D} \right)^2 \qquad (2)$$

$P_t$ the transmitted power
$P_r$ the received power
$G_t$ the transmitting omni basic antenna gain
$G_r$ the receiving antenna gain
$D$ the distance between the two antennas
$\lambda$ the wavelength of the signal

From (2) yields the formula the range of coverage as follows:

$$D = \frac{1}{\frac{4\pi}{\lambda} \sqrt{\frac{P_r}{P_t G_t G_r}}} \qquad (3)$$

We note that as the frequency increases, the range decreases. The figure 3 shows the variation of signal range based on the transmission frequency for a fixed power. The most revealing characteristic of this graph is the non-linearity. The signals of GSM/GPRS with 900 MHz propagate much better than ZigBee, Wi-Fi, Bluetooth with 2.4 GHz and UWB with 3.1 GHz vice to vice coverage area.

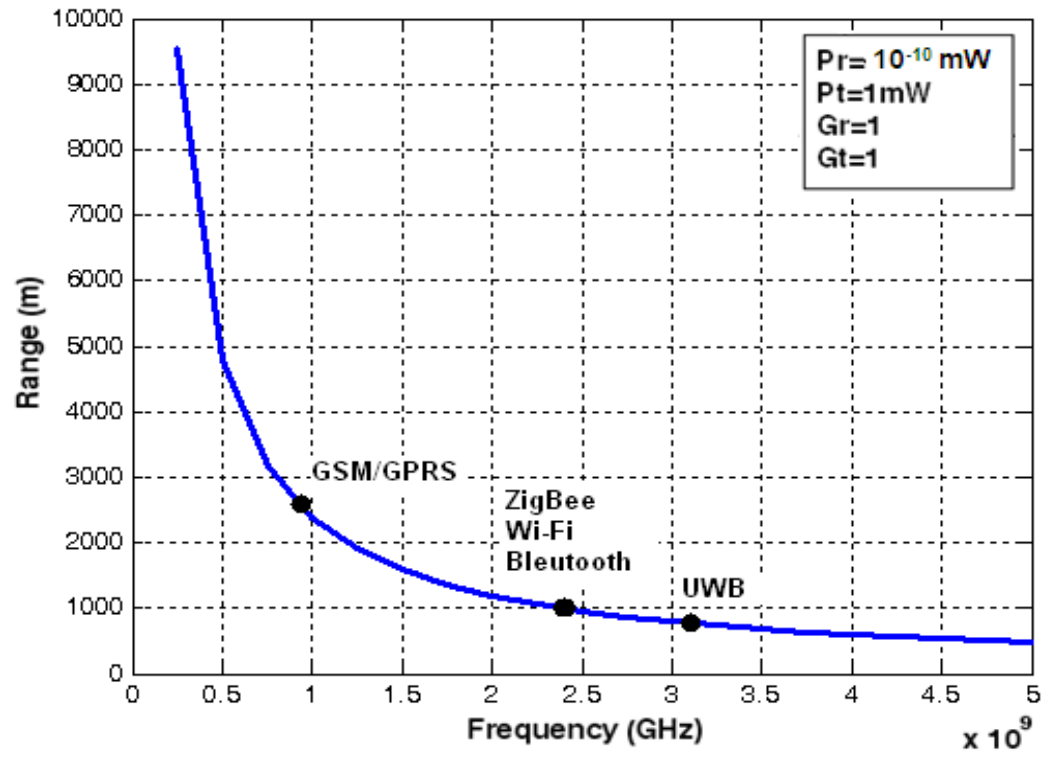

Fig. 3. Range evolution according to the transmission frequency

### D. *Energy consumption*

The energy consumption for intelligent sensor involves three steps: acquisition, communication, computation and data aggregation. This consumption in the acquisition operation depends on the nature of the application [3]. Data traffic, particularly in the transmission, consumes more energy than the other operations. It also depends on the distance between the transmitter and receiver [4], [5]. According to the radio energy model, [6], [38-44] the transmission power of a k bit message to a distance d is given by:

$$E_{TX}(k,d)=\begin{cases} k.\varepsilon_{fs}.d^2+k.E_{Elec} & d<d_0 \\ k.\varepsilon_{amp}.d^4+k.E_{Elec} & d\geq d_0 \end{cases} \quad (4)$$

$$d_0=\sqrt{\frac{\varepsilon_{fs}}{\varepsilon_{amp}}} \quad (5)$$

$E_{Elec}$ electronic energy
$\varepsilon_{fs}$, $\varepsilon_{amp}$ amplification energy

The electronic energy $E_{Elec}$ depends on several factors such as digital coding, modulation, filtering, and signal propagation, while the amplifier energy depends on the distance to the receiver and the acceptable bit error rate. If the message size and the range of communication are fixed, the required energy to cover a given distance increases.

Table 4. The simulation parameters

| Parameters | Value |
|---|---|
| $E_{Elec}$ | 50 nJ/bit |
| $\varepsilon_{fs}$ | 10 pJ/bit/m$^2$ |
| $\varepsilon_{amp}$ | 0.0013 pJ/bit/m$^4$ |

The figure 4 illustrates the evolution of the energy consumption for ZigBee protocol based on the signal range. We can say that an increase in data packet size allows then an increase of the transmission energy. Equations (4) and (5) can be generalized for the all wireless mentioned protocols. The simulation parameters are given in table 4.

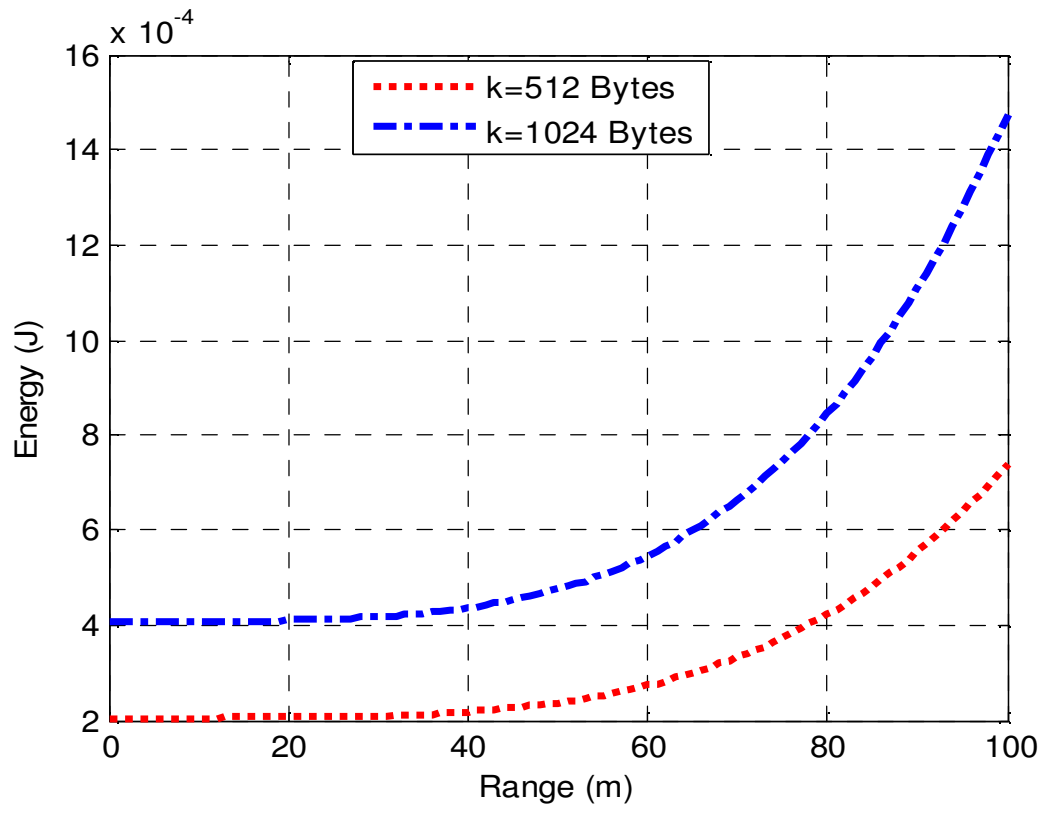

Fig. 4. The energy consumption depending on the signal range

The predicted received power by an intelligent sensor for each data packet according to the communication range d is given by the Two-Ray Ground and the Friss free space models [3], [35], [40] as follows:

$$P_r(d)=\begin{cases} \frac{P_tG_tG_r\lambda^2}{(4\pi d)^2 L} & d<d_c \\ \frac{P_tG_tG_rh_t^2h_r^2}{d^4} & d\geq d_c \end{cases} \quad (6)$$

$$d_c=\frac{4\pi\sqrt{L}h_rh_t}{\lambda} \quad (7)$$

$L$ the path loss
$h_t$ the height of the transmitter antenna
$h_r$ the height of the receiver antenna
$d$ the distance between transmitter and receiver

The figure 5 shows the evolution of the reception power based on the signal range for different studied protocols and for fixed data packet size:

Table 5. The simulation parameters

| Parameters | Value |
|---|---|
| L | 1 |
| $G_t$=$G_r$ | 1 |
| $h_t$=$h_r$ | 1.5 m |

| Protocols | Transmitted Power (Watt) |
|---|---|
| Bluetooth | 0.1 |
| UWB | 0.04 |
| ZigBee | 0.0063 |
| Wi-Fi | 1 |
| Wi-Max | 0.25 |
| GSM/GPRS | 2 |

According to this figure, it is noted that when the distance between the transmitter and the receiver increases, the received power decreases, this is justified by the power loss in the path. The ZigBee, UWB and Bluetooth have low power consumption while Wi-Max, Wi-Fi and GPRS absorb more power due to theirs high communication range reason.

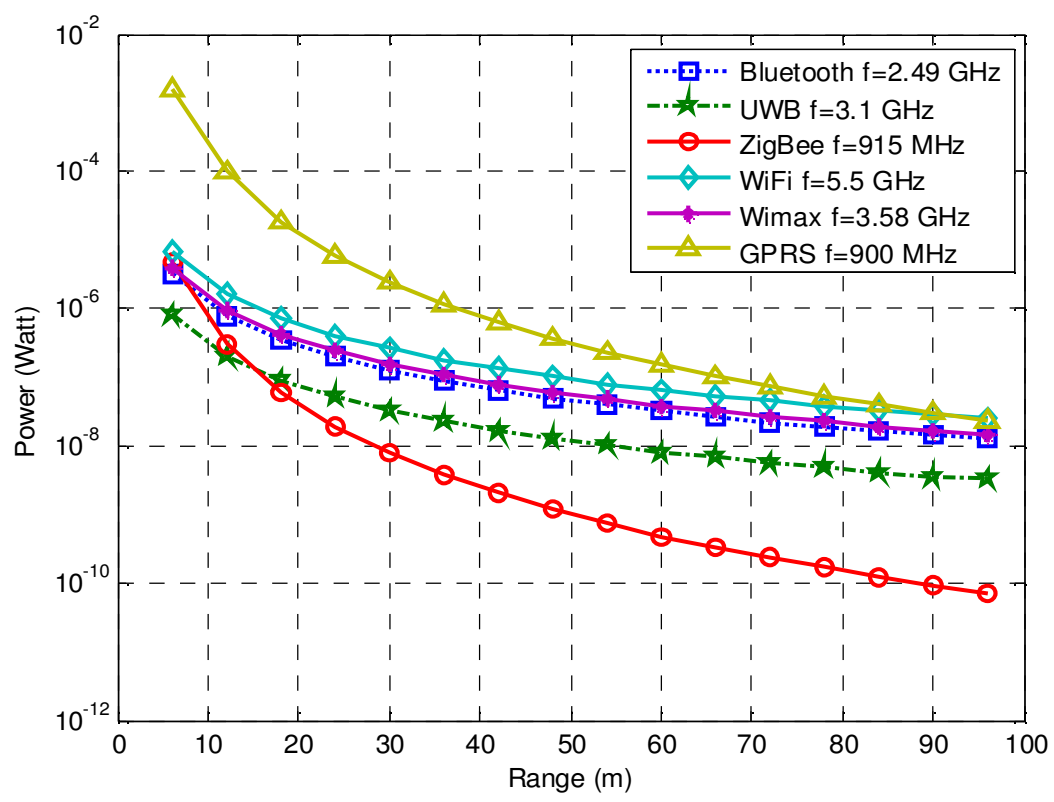

Fig. 5. The received power depending on the signal range with fixed message size

### E. *Chipset power consumption*

To compare practically the power consumption, we are presents in the table 6 the detailed representative characteristics of particular chipset for each protocol [44-49]. The figure 6 shows the consumption power in (mW) for each protocol. Obviously we note that Bluetooth and ZigBee consume less power compared to UWB, Wi-Fi, Wi-Max and a GPRS connection. The difference between the transmission power and reception power for the protocols GPRS and Wi-Max is justified by the power loss due to the attenuation of the signal in the communication path since both of these protocols have a large coverage area.

Table 6. Power consumption characteristics of chipsets

| Protocols | Chipset | $V_{DD}$ (volt) | $I_{TX}$ (mA) | $I_{RX}$ (mA) | Bit rate (Mb/s) |
|---|---|---|---|---|---|
| Bluetooth | BlueCore2 | 1.8 | 57 | 47 | 0.72 |
| UWB | XS110 | 3.3 | ~227 | ~227 | 114 |
| ZigBee | CC2430 | 3.0 | 24.7 | 27 | 0.25 |
| Wi-Fi | CX53111 | 3.3 | 219 | 215 | 54 |
| Wi-Max | AT86 RF535A | 3.3 | 320 | 200 | 70 |
| GPRS | SIM300 | 3 | 350* | 230* | 0.164* |

* For GSM 900 DATA mode, GPRS ( 1 Rx,1 Tx )

Based on the data rate of each protocol, the normalized energy consumption in (mJ/Mb) is shown in the figure 7 for a data size equal to 1 Mb. This shows clearly in this figure that the UWB, Wi-Fi and Wi-Max have better energy efficiency.

In summary, we can say that Bluetooth and ZigBee are suitable for low data rate applications with a limited battery power, because of their low energy consumption which promotes a long lifetime.

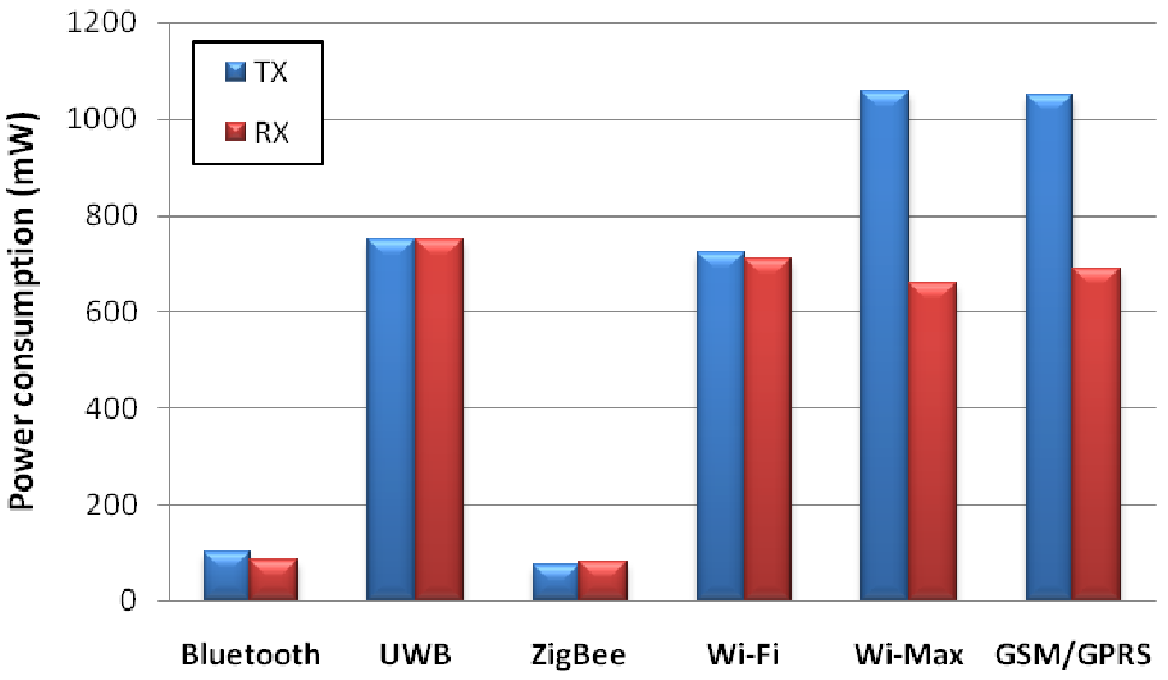

Fig. 6. Comparison of chipset power consumption for each protocol

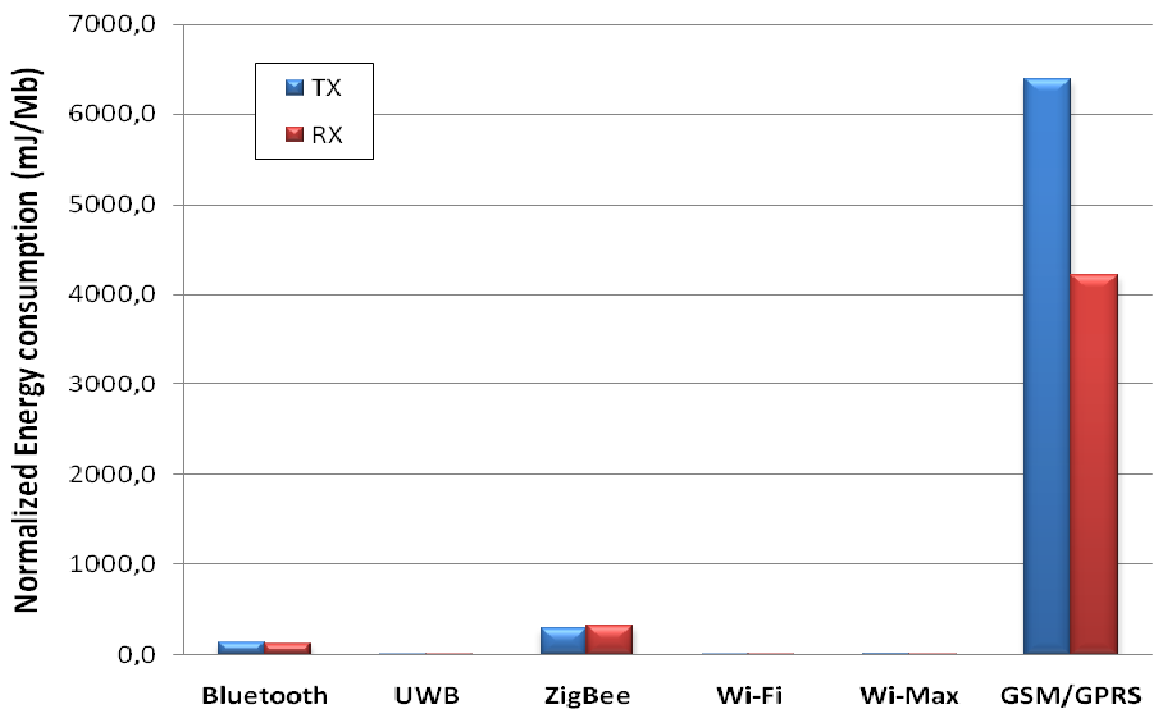

Fig. 7. Comparing the chipset normalized energy consumption for each protocol

Contrariwise for implementations of high data rate, UWB, Wi-Fi and Wi-Max would be the best solution due to their low normalized energy consumption. While for monitoring and surveillance applications with low data rate requiring large area coverage, GPRS would be an adequate solution.

### F. *Bit error rate*

The transmitted signal is corrupted by white noise AWGN (Additive White Gaussian Noise) to measure the performance of the digital transmissions (OQ-B-Q-PSK, 4PAM, 16QAM, GMSK, GFSK, 8DPSK, 8PSK and OFDM), used in the studied wireless protocols, by calculating the bit error probability. This latter is a very good tool to measure the performance of a modulation used in a communication module and therefore helps to improve its robustness. It is calculated by the following formula:

$$\mathrm{BER} = \frac{N_{Err}}{N_{TXBits}} \quad (8)$$

$N_{Err}$ the number of errors

$N_{TXBits}$ the number of transmitted bits

The figure 8 shows the BER of the different modulations used in wireless technologies mentioned above according on signal to noise ratio $E_b/N_0$. The BER for all systems decreases monotonically with increasing values of $E_b/N_0$, the curves defining a shape similar to the shape of a waterfall [36], [38]. The BER for QPSK and OQPSK is the same as for BPSK. We note that the higher order modulations exhibit higher error rates which thus leads to a compromise with the spectral efficiency. QPSK and GMSK seem the best compromise between spectral efficiency and BER followed by other modulations. These two robust modulations are used in Wi-MAX, ZigBee, Wi-Fi and in GPRS network. They can be employed in the noisy channels and in the noisy environments.

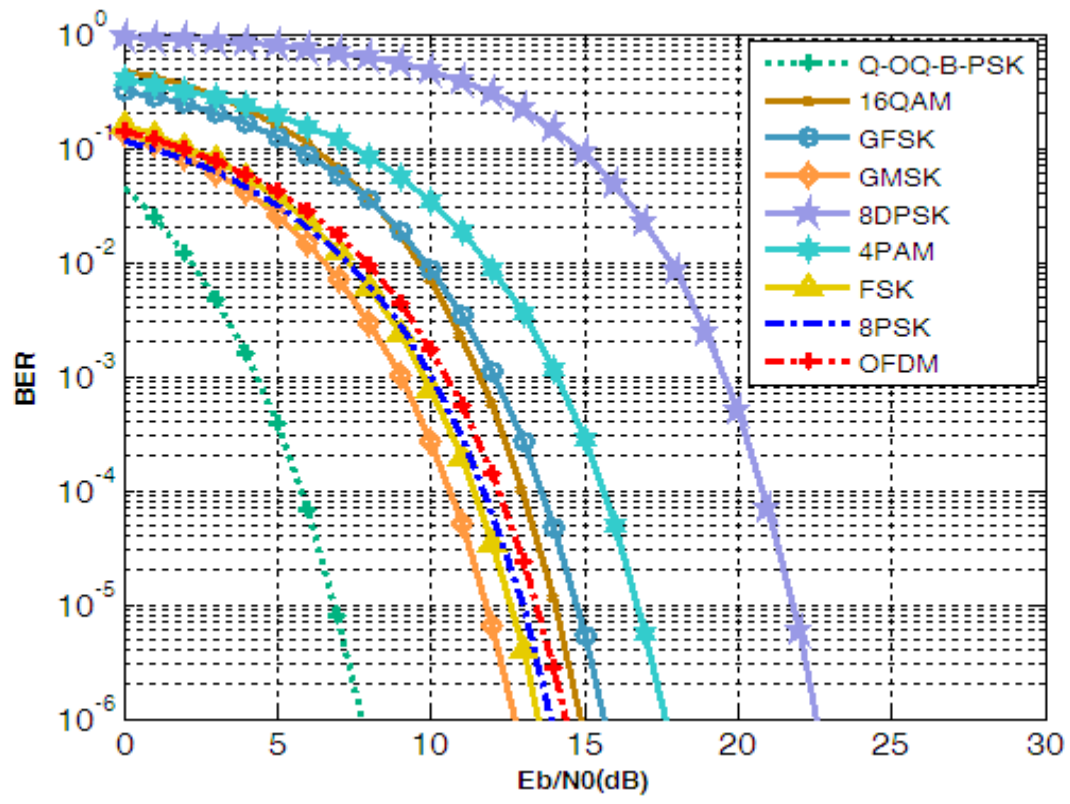

Fig. 8. Bit Error Rate for different modulations

Table 7. $E_b/N_0$ values which cancels BER for the different modulations

| Modulation | $E_b/N_0$ (dB) | B.E.R |
|---|---|---|
| B-OQ-QPSK | 7,8 | $10^{-6}$ |
| GMSK | 12,7 | $10^{-6}$ |
| FSK | 13,3 | $10^{-6}$ |
| 8PSK | 13,8 | $10^{-6}$ |
| OFDM | 14,3 | $10^{-6}$ |
| 16QAM | 14,8 | $10^{-6}$ |
| GFSK | 15,7 | $10^{-6}$ |
| 4PAM | 17,6 | $10^{-6}$ |
| 8DPSK | 22,6 | $10^{-6}$ |

However, because of their sensitivity to noise and non-linearities, the modulations 4PAM and 8DPSK remain little used compared to other modulations. Concerning the QAM modulation, it uses more efficiently the transmitted energy when the number of bits per symbol increases. As for the frequency hopping FSK modulations, the increase of the symbols will enable reduction of the BER but also increase the spectral occupancy. The main fault of these FSK modulations is their low spectral efficiency. On the other side, the GMSK modulation has been developed in order to increase the spectral efficiency [50]. It has a satisfactory performance in terms of BER and noise resistance [9], [35-41]. Furthermore, the lower bit error probability is obtained to the detriment of the number of users. It can be helpful to investigate the relationship between the transmission quality and the number of served users.

### G. *Data coding efficiency*

The coding efficiency can be calculated from the following formula:

$$P_{cdeff} = 100 \times \frac{N_{data}}{\left( N_{data} + \left( \frac{N_{data}}{N_{maxPld}} \right) \times N_{ovhd} \right)} \quad (9)$$

Based on the figure 9, the coding efficiency increases when the data size increase. For small data size, Bluetooth and ZigBee are the best solutions while for high data sizes GPRS, UWB, Wi-Max and Wi-Fi have efficiency around 94%. For point of view application, the automation industrial systems based on intelligent sensors, since most data monitoring and industrial control have generally a small size because they don't require an important data rate such as pressure or temperature measurements which don't exceeds 4 bytes, Bluetooth, ZigBee and GPRS can be a good choice due to their coding efficiency and their low data rate. On the other hand, for applications requiring a large cover zone as the borders monitoring, the persons tracking or the environmental monitoring or the event detection, GPRS and Wi-Max are an adequate solution, whereas for the multimedia applications requiring an important data rate such the video monitoring, Wi-Fi, UWB and Wi-Max form a better solution.

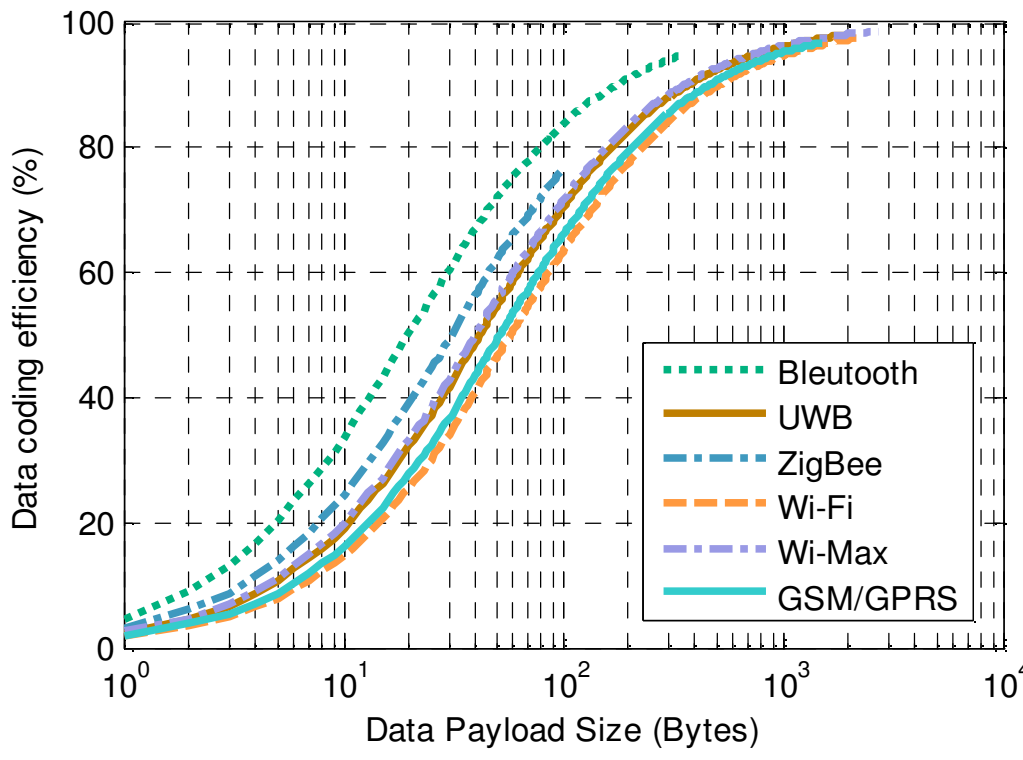

Fig. 9. Coding efficiency depending on data size

### H. *Packet error probability*

The packet error probability is one indicator which can be useful to know QoS level, if packet error is detected an error-correction mechanism is activated by retransmission of the faulty packet at the MAC layer. This process improves system reliability, but increases delay. It can be calculated as follows [52]:

$$p_e = 1 - (1 - b_e)^L \quad (10)$$

$b_e$ bit error probability
L packet length in bits

Referring in figure 10, it is obviously clear that increasing L may lead to higher packet error probability and therefore a higher number of retransmissions. Depending on the bit-error probability and packet overhead, packet length L can be optimized so that an optimal system lifetime can be achieved.

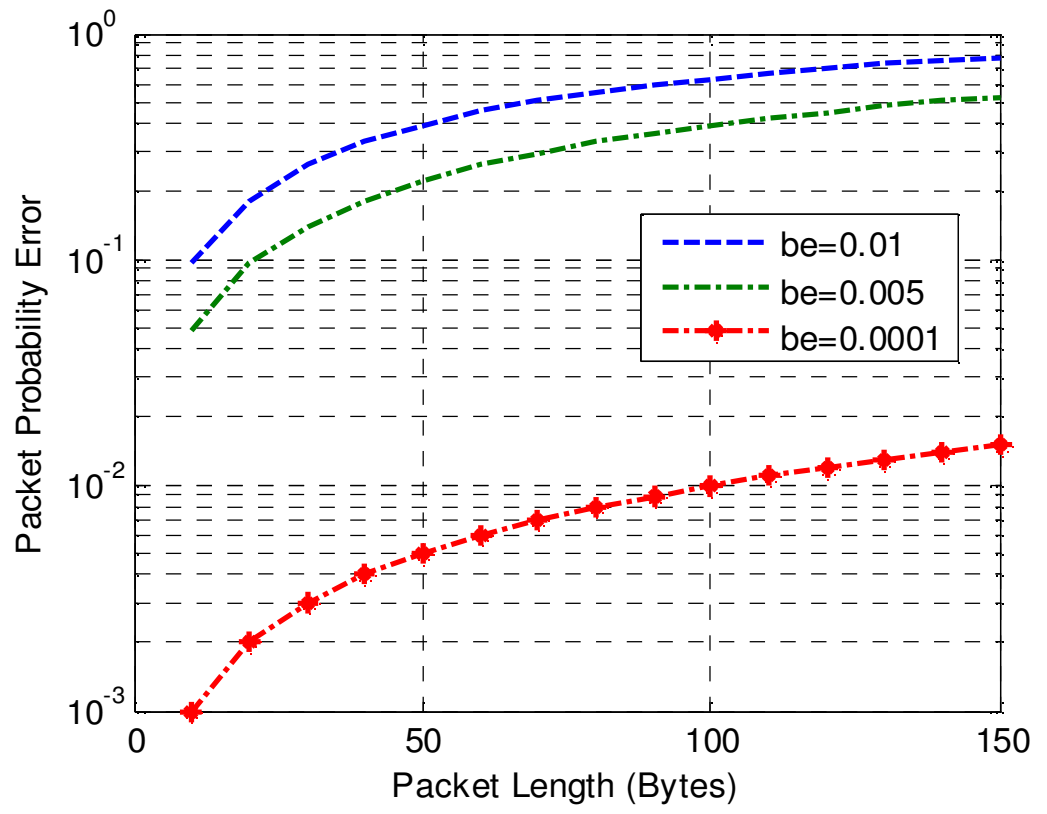

Fig. 10. Packet error depending on packet length

### *I. Energy index*

Energy index is used to measure the transmission energy cost of a sensor node and therefore to determine life span of the sensors network. It is given as follows [52]:

$$E_i = \frac{L-O}{(1+n_r).(e_t+e_m+e_c)} \quad (11)$$

$n_r$ number of retransmissions
O packet overhead
$e_t$ transceiver energy
$e_m$ collision, idle and overhearing energy

Figure 11 presents numerical results of $E_i$ as a function of packet length when:

O=2 bytes    $e_m$ =200 nJ
$e_t$ =100L nJ    $e_c$ =100 nJ

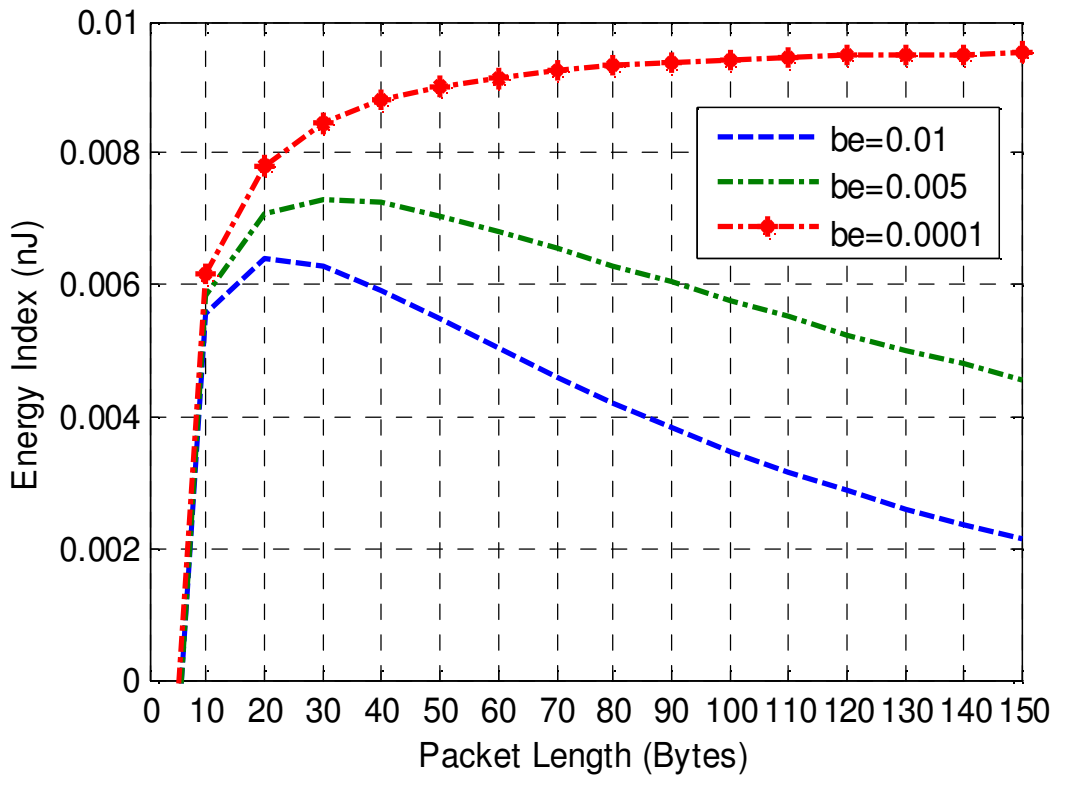

Fig. 11. Energy index versus packet length

It can be observed that in fact there is an optimal value of packet length L which maximizes $E_i$. The optimal value of L increases with a lower bit error rate and/or with an increase in the packet overhead O. When designing and deploying a wireless sensor network, it is possible to choose the appropriate values of the studied parameters that will extend system lifetime.

### *J. Real time throughput*

The real time throughput is defined to know the number of octets per second transmitted on a specific link exclusively relevant to real-time traffic [52]. It analytical expression relevant to the link between the devices is the following:

$$\text{Th} = \frac{m}{T_{frame} + T_{backoff}} \quad (12)$$

m the amount of data to be transmitted
$T_{frame}$ time required to transmit a MAC data frame
$T_{backoff}$ the average backoff time

Figure 12 shows the simulation results for two values of transmitted data m=512bytes and m=1024bytes with a fixed sensor node bandwidth equal to 38,4Kbits/s and with $T_{frame}$=11,39ms, It seems that real time throughput needed to convey data decrease when transmitted data size decrease. The behavior of Th has a high bound and decrease suddenly for the lowest value of $T_{backoff}$, thereafter it reduce asymptotically with growing values of $T_{backoff}$. This fall which occurs in throughput is due to the bandwidth waste caused mainly by collisions and back-off delays when two or more sensor nodes attempt to transmit measurements at the same time which can produce saturation phenomena.

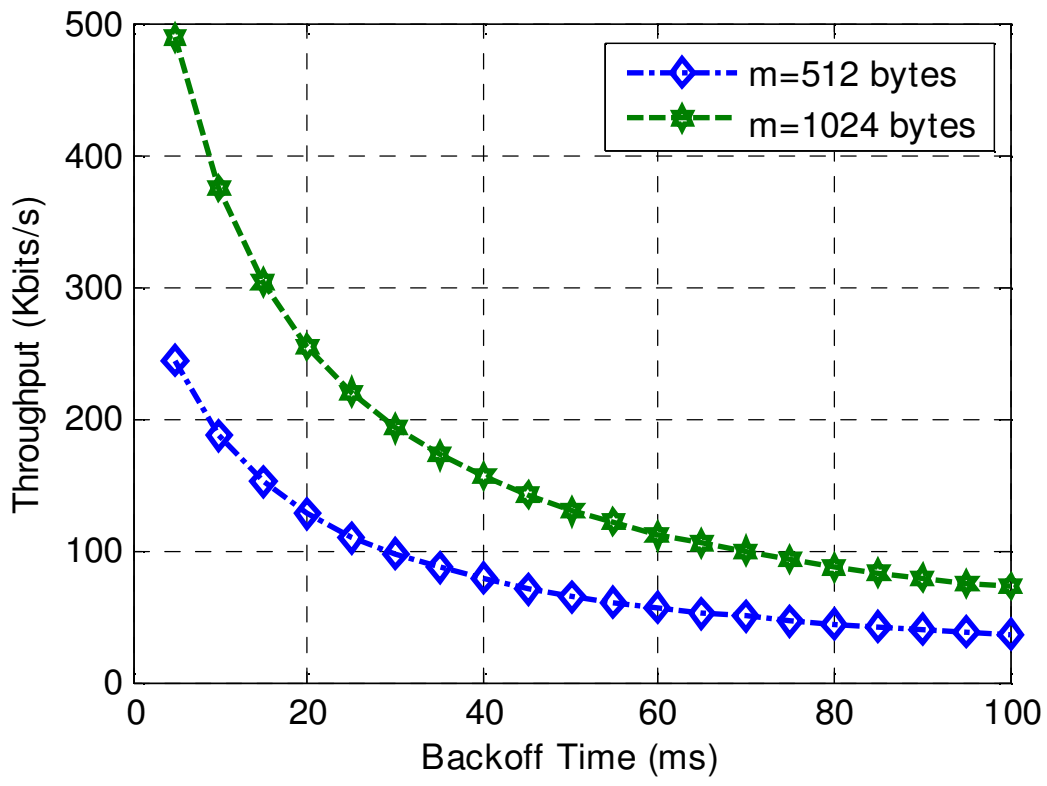

Fig. 12. Throughput versus backoff time

### *K. Microcontrollers energy efficiency*

Energy efficient computing during the sensing algorithm process depends on the number of instructions executed by the software implemented into microcontroller of the sensor motes [53]. This energy is the sum of energy required to switch between the internal states of the MCU given by (14) and the leakage energy which refers to the energy lost while the MCU is idle modeled by (15).

$$E_{total} = E_{switch} + E_{leakage} \quad (13)$$

$$E_{switch} = C_{total} V_{dd}^2 \quad (14)$$

$$E_{leakage} = V_{dd}\left(I_0 e^{\frac{V_{dd}}{nV_T}}\right)\left(\frac{N}{f}\right) \quad (15)$$

Where N is the number of cycles which the program takes to be executed, $f$ is the MCU clock frequency and $V_T$ is the thermal voltage. Figure 13 shows comparison between the behaviors of the energy consumed per instruction in (J) depending on the number of instruction cycles for types of sensor motes. The energy consumption increases with the augmentation of the number of instruction cycles. This consumption is very important for an MCU working with lowest clock frequency. MCU leakage energy is an important parameter to optimizing when designing a wireless microsensor mote because energy is wasted while no work is done.

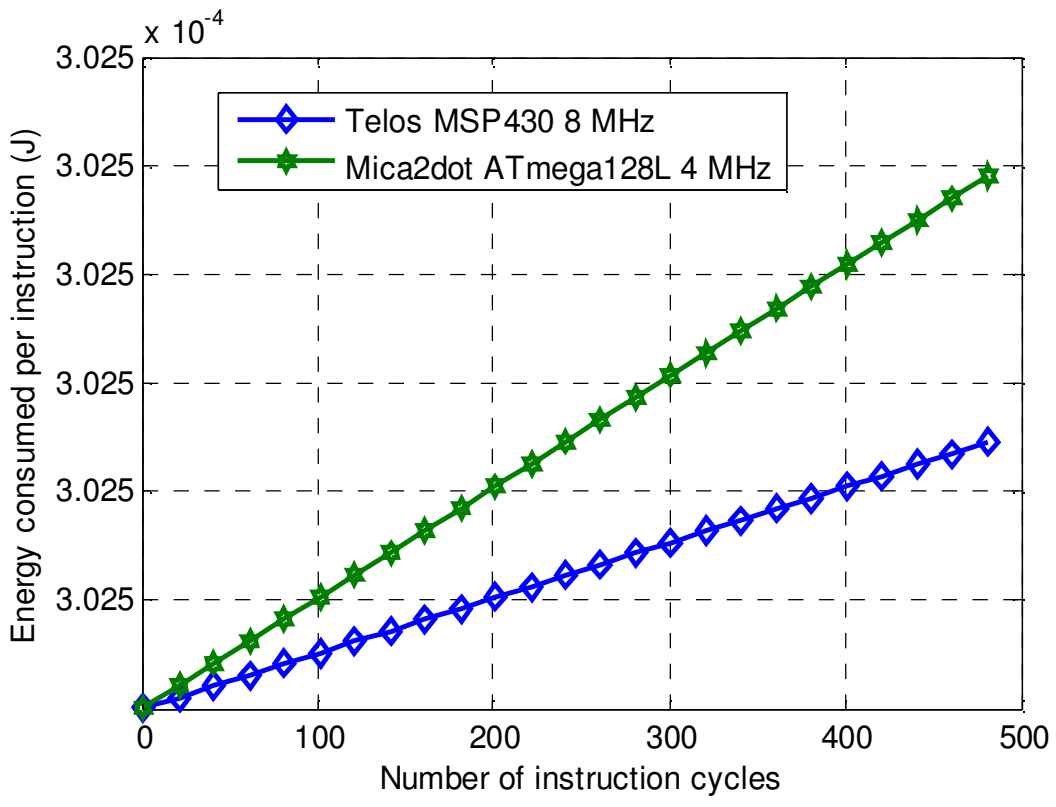

Fig. 13. Computation energy comparison

## V. SENSORS MOTES COMPARAISON

While the particular sensor types vary significantly depending on the application, a variety of wireless modules and hardware platforms have been realized to facilitate developing applications in intelligent sensors networks [51]. Figure 14 summarizes a comparison of popular sensor motes that were designed in the recent few years. As can be observed, the capabilities of these platforms vary significantly. However, this will assist researchers, developers and users to select the most appropriate platform for their purposes.

## VI. CONCLUSION

We have presented in this paper a comparative performance analysis of six wireless protocols: Bluetooth, UWB, ZigBee, Wi-Fi, Wi-Max and GSM/GPRS. Choosing a wireless protocol for an intelligent sensing application must do compromise between the energy cost, QoS and real time execution. However, robust sensing quantitative evaluation indicators permitted us to determine the suitable protocol for an application based on intelligent sensor. Furthermore, the adequacy of the sensor networks is influenced strongly by many others factors as the network reliability, the link capacity between several networks having different communication protocols, the security, the chipset price, conformity and installation cost. The challenge is to develop a gateway (multi-standard transceiver) that enables data exchange between these heterogeneous infrastructures with a good QoS. As perspective of this work, it is proposed to extend this study to investigate the impact of modulation scheme on the sensing coverage, on the bandwidth, on the number of served users and on energy consumption.

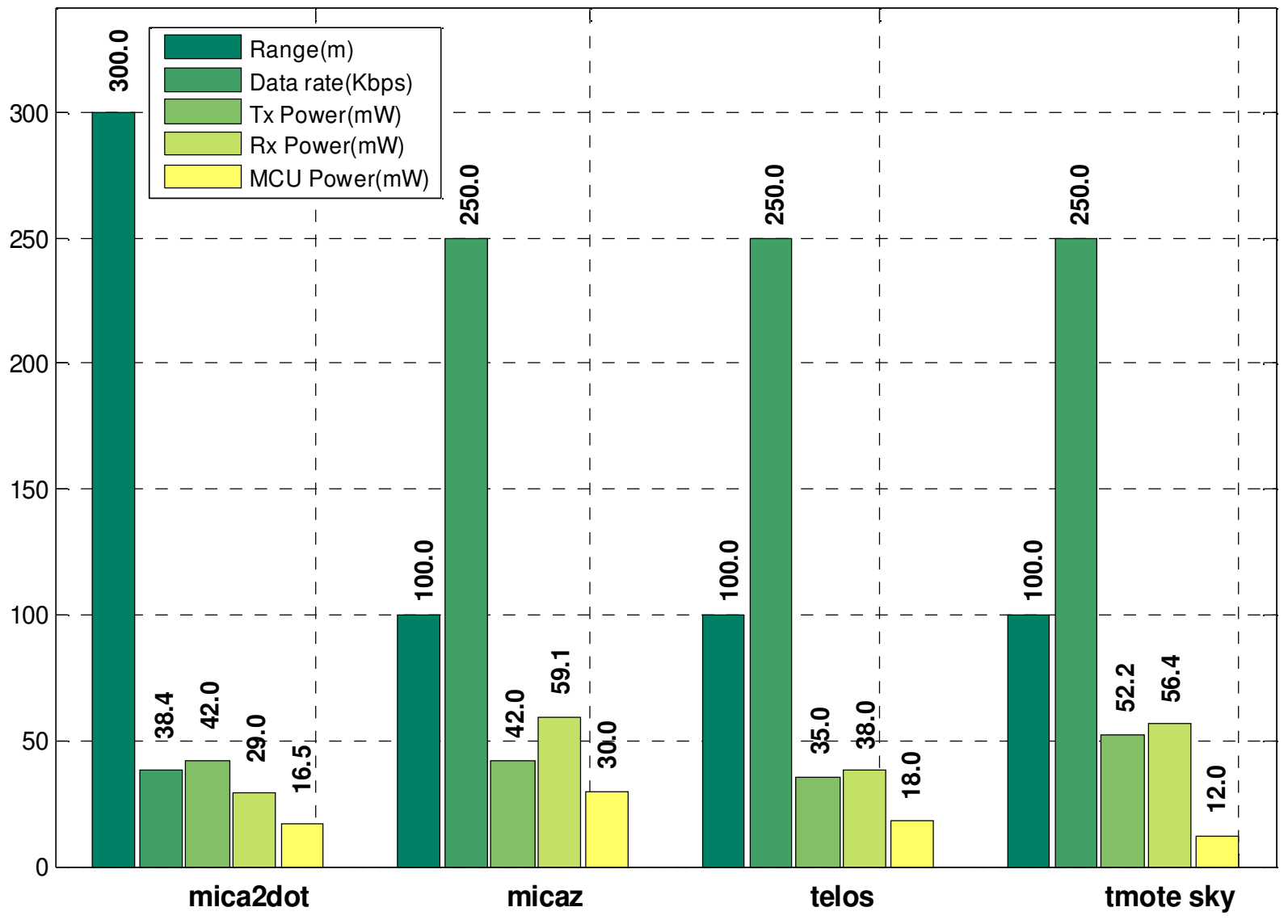

Fig. 14. Popular sensor motes comparison

## REFERENCES

[1] Jean-Paul M., G. Linmartz's, "Wireless Communication, The Interactive Multimedia CD-ROM", Baltzer Science Publishers, Amsterdam, Vol. 1 (1996), No.1

[2] Helen Fornazier et al., "Wireless Communication : Wi-Fi, Bluetooth, IEEE 802.15.4, DASH7", ROSE 2012 ELECINF344 / ELECINF381, Télécom ParisTech, web site : http://rose.eu.org/2012/category/admin

[3] Lehsaini Mohamed, "Diffusion et couverture basées sur le clustering dans les réseaux de capteurs : application à la domotique ", Thèse de Doctorat, Université de Franche-Comté Besançon, U.F.R Sciences et Techniques, École Doctorale SPIM, Juillet 2009

[4] Trevor Pering et al., CoolSpots: "Reducing the Power Consumption of Wireless Mobile Devices with Multiple Radio Interfaces", ACM 1-59593-195-3/06/0006, MobiSys'06, June 19–22, 2006, Uppsala, Sweden

[5] Travis Collins et al., "A Green Approach to a Multi-Protocol Wireless Communications Network", Major Qualifying Project to obtain the Degree in Bachelor of Science in Electrical and Computer Engineering, Faculty of Worcester Polytechnic Institute, University of Limerick 2011, http://www.wpi.edu/Academics/Projects

[6] Wendi B. Heinzelman et al., "An Application-Specific Protocol Architecture for Wireless Microsensor Networks", IEEE TRANSACTIONS ON WIRELESS COMMUNICATIONS, VOL. 1, NO. 4, OCTOBER 2002

[7] Baghouri Mostafa, Chakkor Saad, Hajraoui Abderrahmane, "Fuzzy logic approach to improving Stable Election Protocol for clustered heterogeneous wireless sensor networks", Journal of Theoretical and Applied Information Technology, Vol. 53 No.3, July 2013

[8] El Ahmadi Cheikh, Chakkor Saad et al, "New Approach to Improving Lifetime in Heterogeneous Wireless Sensor Networks Based on Clustering Energy Efficiency Algorithm", Journal of Theoretical and Applied Information Technology, Vol. 61 No.2, March 2014

[9] Crepin Nsiala Nzeza, " RÉCEPTEUR ADAPTATIF MULTI-STANDARDS POUR LES SIGNAUX A ÉTALEMENT DE SPECTRE EN CONTEXTE NON COOPÉRATIF ", thèse de Doctorat, UNIVERSITÉ de Bretagne Occidentale, Juillet 2006

[10] Guillaume Terrasson, "CONTRIBUTION A LA CONCEPTION D'EMETTEUR- RECEPTEUR POUR MICROCAPTEURS AUTONOMES", thèse de Doctorat, UNIVERSITÉ BORDEAUX 1, Novembre 2008

[11] Creed Huddleston, "Intelligent Sensor Design Using the Microchip dsPIC (Embedded Technology)", Newnes 2007

[12] Elmostafa Ziani, "CONCEPTION ET REALISATION D'UN INSTRUMENT ULTRASONORE INTELLIGENT DEDIE A LA MESURE DE DEBITS D'ECOULEMENT A SURFACE LIBRE", Thèse de doctorat en cotutelle, Université Abdelmalek Essaâdi, Faculté des sciences et techniques de Tanger, Université de paris 13 Villetaneuse, Institut universitaire de technologie de Saint- denis 2005

[13] Fei Hu, Qi Hao, "Intelligent Sensor Networks: The Integration of Sensor Networks", Signal Processing and Machine Learning, CRC Press 2012

[14] Jin-Shyan Lee et al., "A Comparative Study of Wireless Protocols: Bluetooth, UWB, ZigBee", and Wi-Fi, The 33rd Annual Conference of the IEEE Industrial Electronics Society (IECON), Taipei, Taiwan, November 5-8, 2007

[15] Adil Koukab et al., "A GSM-GPRS/UMTS FDD-TDD/WLAN 802.11a-b-g Multi-Standard Carrier Generation System", IEEE JOURNAL OF SOLID-STATE CIRCUITS, VOL. 41, NO. 7, JULY 2006

[16] Jin-Shyan Lee, "Performance Evaluation of IEEE 802.15.4 for Low-Rate Wireless Personal Area Networks", IEEE Transactions on Consumer Electronics, Vol. 52, No. 3, AUGUST 2006

[17] Klaus Gravogl et al., "Choosing the best wireless protocol for typical applications", 2nd Workshop on Ultra-low Power Wireless Sensor Networks (WUPS 2011),Como, Italy, February 2011

[18] Z. Mammeri, "Réseaux sans fils Caractéristiques et principaux standards", M1 Info Cours de Réseaux, IRIT, Université Paul Sabatier, Toulouse http://www.irit.fr/~Zoubir.Mammeri/Chap6WLAN.pdf

[19] Ghobad Heidari, "WiMedia UWB: Technology of Choice for Wireless USB and Bluetooth", edition John Wiley & Sons Ltd 2008

[20] Ms. Dharmistha, D. Vishwakarma, "IEEE 802.15.4 and ZigBee: A Conceptual Study", International Journal of Advanced Research in Computer and Communication Engineering, ISSN : 2278 – 1021, Vol. 1, Issue 7, September 2012

[21] Vaddina Prakash Rao, "The simulative Investigation of Zigbee/IEEE 802.15.4", Master Thesis of Science, DRESDEN UNIVERSITY OF TECHNOLOGY, FACULTY OF ELECTRICAL ENGINEERING AND INFORMATION TECHNOLOGY, Department of Electrical Engineering and Information Technology, Chair of Telecommunications, September, 2005

[22] http://www.zigbee.org/Specifications/ZigBeeIP/Overview.aspx

[23] Reen-Cheng Wang et al., "Internetworking Between ZigBee/802.15.4 and IPv6/802.3 Network", ACM 978-1-59593-790-2/07/0008, IPv6'07, August 31, 2007, Kyoto, Japan

[24] Aurélien Géron, "WIFI PROFESSIONNEL La norme 802.11, le déploiement, la sécurité", 3ème édition DUNOD

[25] Bhavneet Sidhu et al., "Emerging Wireless Standards - WiFi, ZigBee and WiMAX", World Academy of Science, Engineering and Technology 2007

[26] Michèle Germain, "WiMAX à l'usage des communications haut débit", Forum atena, lulu.com, Paris, 2009

[27] Loutfi Nuaymi, "WiMAX : Technology for Broadband Wireless Access", Wiley 2007

[28] Marwa Ibrahim et al., "Performance investigation of Wi-Max 802.16m in mobile high altitude platforms", Journal of Theoretical and Applied Information Technology, 10th June 2013. Vol. 52 No.1

[29] Xavier Lagrange et al., " Réseaux GSM : des principes à la norme ", Éditions Hermès Sciences 2000

[30] Timo Halonen et al., "GSM, GPRS Performance and EDGE, Evolution Towards 3G/UMTS", Second Edition, John Wiley & Sons Ltd 2003

[31] Brahim Ghribi, Luigi Logrippo, "Understanding GPRS: the GSM packet radio service", Elsevier Computer Networks 2000, pages 763 - 779

[32] Christian Bettstetter et al., "GSM phase 2+general packet radio service GPRS: architecture, protocols, and air interface", IEEE Communications Surveys, Third Quarter 1999, vol. 2 no. 3, http://www.comsoc.org/pubs/surveys

[33] Joseph Ho et al., "Throughput and Buffer Analysis for GSM General Packet Radio Service (GPRS)", Wireless

Communications and Networking Conference New Orleans, LA, 1999. WCNC. 1999 IEEE, Pages 1427 - 1431 vol.3

[34] Constantine A. Balanis, "Antenna Theory: Analysis and Design",2nd edition John Wiley and Sons, Inc 1997

[35] François de Dieuleveult, Olivier Romain, "ÉLECTRONIQUE APPLIQUÉE AUX HAUTES FRÉQUENCES" Principes et applications, 2e édition, Dunod, Paris 2008

[36] DR. Kamilo Feher, "Wireless Digital Communications (Modulation & Spread spectrum Applications)", PHI Learning, Prentice Hall PTR, 1995

[37] Simon Haykin, "Communication Systems", 4Th Edition with Solutions Manual, John Wiley and Sons, Inc 2001.

[38] Proakis, J. G., "Digital Communications", 3rd edition, New York, McGraw-Hill, 1995.

[39] Sklar, B., "Digital Communications: Fundamentals and Applications", Englewood Cliffs, NJ, Prentice-Hall, 1988

[40] Rappaport T.S., "Wireless Communications Principles and Practice", 2nd Edition, Prentice Hall, 2001

[41] Andreas F.Molisch, "WIRELESS COMMUNICATIONS", John Wiley & Sons Ltd., Second Edition 2011

[42] Prakash C. Gupta, "Data Communications and Computer Networks", PHI Learning, Prentice-Hall of India 2006

[43] Nitin Mittal et al., "IMPROVED LEACH COMMUNICATION PROTOCOL FOR WSN", NCCI 19-20 March 2010, National Conference on Computational Instrumentation CSIO Chandigarh, INDIA

[44] Cambridge Silicon Radio, BlueCore2-External Product Data Sheet. Cambridge, UK, Aug. 2006

[45] Freescale, XS110 UWB Solution for Media-Rich Wireless Applications. San Diego, CA, Dec. 2004

[46] Chipcon, CC2430 Preliminary Data Sheet (rev. 1.03). Oslo, Norway, 2006

[47] Conexant, Single-Chip WLAN Radio CX53111. Newport Beach, CA, 2006

[48] SIMCOM Ltd, SIM300 Hardware Specification, 27th Dec 2005

[49] ATMEL, WiMax Transceiver 802.16-2004, AT86RF535A Preliminary Data Sheet, 2006

[50] David K. Asano, Subbarayan Pasupathy, "Optimization of Coded GMSK Systems", IEEE Transactions on Information Theory, VOL. 48, NO. 10, OCTOBER 2002

[51] Kazem Sohraby et al., "Wireless Sensor Networks Technology, Protocols, and Applications", John Wiley & Sons, Inc. Edition 2007

[52] Giovanni Gamba et al., "Performance Indicators for Wireless Industrial Communication Networks", 8th IEEE International Workshop on Factory Communication Systems WFCS 2010

[53] E. Shih, et al., "Physical Layer Driven Protocol and Algorithm Design for Energy-Efficient Wireless Sensor Networks", 7th Annual ACM SIGMOBILE Conference on Mobile Computing and Networking, Rome, 2001

[54] "Guide of MATLAB" 7.8.0 (R2009a), www.mathworks.com

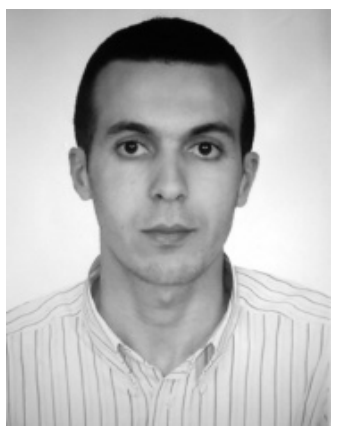

**Chakkor Saad** was born in Tangier Morocco. He's a member in the Physics department, Team Communication and detection Systems, Faculty of sciences, University of Abdelmalek Essaâdi, Tetouan Morocco, and his research area is: wireless intelligent sensors and theirs applications, frequency estimation algorithms for faults detection and diagnosis system in electromecanical machines. He obtained the Master's degree in Electrical and Computer Engineering from the Faculty of Sciences and Techniques of Tangier, Morocco in 2002. He graduated enabling teaching computer science for secondary qualifying school in 2003. In 2006, he graduated from DESA in Automatics and information processing at the same faculty. He works as teacher of computer science in the high school.

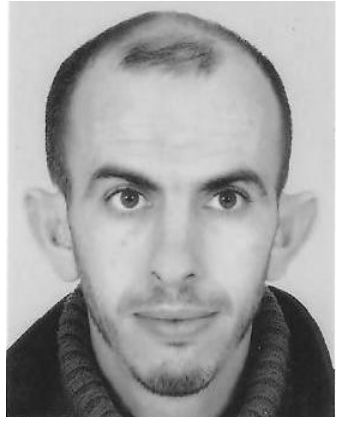

**Baghouri Mostafa** was born in Tangier Morocco. He's a member in the Physics department, Team Communication Systems, Faculty of sciences, University of Abdelmalek Essaâdi, Tetouan Morocco, his research area is: routing and real time protocols for energy optimization in wireless sensors networks. He obtained a Master's degree in Electrical and Computer Engineering from the Faculty of Science and Technology of Tangier in Morocco in 2002. He graduated enabling teaching computer science for secondary qualifying school in 2004. In 2006, he graduated from DESA in Automatics and information processing at the same faculty. He work teacher of computer science in the high school.

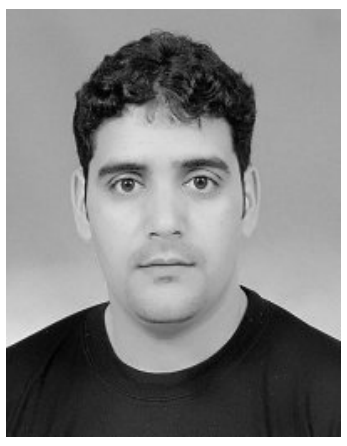

**El ahmadi Cheikh** was born in Boujdour Morocco. He's a member in the Physics department, Team Communication and detection Systems, Faculty of sciences, University of Abdelmalek Essaâdi, Tetouan Morocco, and his research area is: improving performance of sensor networks. He obtained the Master's degree in Networks and Systems from the Faculty of Sciences and Techniques of Settat, Morocco in 2009. His current research interests are in the areas of embedded systems, wireless sensor networks, energy efficiency, body sensor networks, and RFID technology.